# Efficient frequency conversion with geometric phase control in optical metasurfaces

*Bernhard Reineke Matsudo[1], Basudeb Sain[1], Luca Carletti[2], Xue Zhang[3], Wenlong Gao[1], Costantino de Angelis[2], Lingling Huang[3], Thomas Zentgraf[1,\*]*

[1] Department of Physics, Paderborn University, Warburger Straße 100, 33098 Paderborn, Germany

[2] Department of Information Engineering and National Institute of Optics (CNR-INO), University of Brescia, Brescia 25123, Italy

[3] School of Optics and Photonics, Beijing Institute of Technology, 100081, Beijing, China

**Abstract**

Metasurfaces have appeared as a versatile platform for miniaturized functional nonlinear optics due to their design freedom in tailoring wavefronts. The key factor that limits its application in functional devices is the low conversion efficiency. Recently, dielectric metasurfaces governed by either high-quality factor modes (quasi-bound states in the continuum) or Mie modes, enabling strong light-matter interaction, have become a prolific route to achieve high nonlinear efficiency. Here, we demonstrate both numerically and experimentally an effective way of spatial nonlinear phase control by using the Pancharatnam-Berry phase principle with a high third harmonic conversion efficiency of $10^{-4}\frac{1}{W^2}$. We find that the magnetic Mie resonance appears to be the main contributor to the third harmonic response, while the contribution from the quasi-bound states in the continuum is negligible. This is confirmed by a phenomenological model based on coupled anharmonic oscillators. Besides, our metasurface provides experimentally a high diffraction efficiency (80-90%) in both polarization channels. We show a functional application of our approach by experimentally reconstructing an encoded polarization-multiplexed vortex beam array with different topological charges at the third harmonic frequency with high fidelity. Our approach has the potential viability for future on-chip nonlinear signal processing and wavefront control.

---

[\*]Email: thomas.zentgraf@uni-paderborn.de

**Introduction**

Nonlinear metasurfaces provide fine-grained control over the generation and efficient manipulation of new frequencies. Freed from the constraints of classical nonlinear optics[1], nonlinear metasurfaces feature versatile functionalities by manipulating the amplitude, phase, and polarization of the generated light. Past research led to essential applications like higher harmonic generation and wave mixing[2,3], beam shaping, wavefront control at higher harmonics[4], vortex beam generation[5], multiplexed holography[6], and nonlinear chirality[7,8]. Many of these applications were demonstrated with plasmonic or hybrid plasmonic-dielectric metasurfaces that are governed by dipole electric or magnetic modes but yield a low conversion efficiency[9,10] in the visible and near-infrared regime. Concerning practical applications, the low conversion efficiency is a severe bottleneck. However, introducing all-dielectric nanostructures made of high refractive index materials with strong nonlinear coefficients allows us to boost the nonlinear conversion efficiency by several orders and introduce certain functionalities[11–22]. These high index nanostructures support diverse spatial electromagnetic modes to couple the material nonlinear coefficients to the external electromagnetic field. Two important categories of these modes in the context of nonlinear optics are the Mie modes and the bound states in the continuum (BICs). Mie modes are resonances in dielectric resonators that show a strong electromagnetic response, allowing to confine the local electromagnetic field inside a subwavelength nanostructure and therefore enhance the light-matter interaction. Consequently, Mie resonances can achieve high nonlinear conversion efficiencies. On the other hand, dielectric nanoresonators can also support BICs. Compared to the Mie modes, BICs feature an infinite quality factor because they are modes decoupled from the free space radiation spectrum[23,24]. By perturbing the symmetry of the BICs, it is possible to transform the BICs into a quasi-BICs (QBICs) capable of coupling to the free-space modes[24–26], which further improves the light-matter interaction and leads to high nonlinear conversion efficiencies.[26,27] Note that design of nonlinear metasurfaces often target resonances at the fundamental wavelength to increase the nonlinear conversion efficiency. However, dual resonance techniques are also useful, where one resonance is placed near the fundamental wavelength and another at the generated harmonic wavelength.[28,29]

Besides addressing the high nonlinear conversion efficiency, nonlinear wavefront control is another important criterion for functional applications. The most common approach for nonlinear phase-tailoring in metasurfaces is the extension of the Huygens principle to the nonlinear regime. Here, the phase of the light is manipulated by tuning the shape of a nanoresonator.[13,30] With this approach, to achieve a fine-grained phase resolution, a set of numerous different nanostructures with distinct phase responses need to be designed, which is a complex task and requires heavy computational resources. A more straightforward approach to achieve a fine-grained phase resolution is the geometric phase or Pancharatnam-Berry (PB) principle. Here, the phase of the generated light is adjusted continuously by using the degree of rotation of a single nanostructure. This approach requires designing and optimizing only a single nanostructure instead of a set of nanostructures for obtaining a high nonlinear conversion efficiency. However, to achieve a high diffraction efficiency and smooth spatial phases, the unit cell size of the metasurface needs to be small. Adversely, a small unit cell size leads to near-field coupling between the nanostructures, which may alter the phase response of the metasurface. Therefore, it is highly challenging to maintain the nonlinear phase relationship resulting from the geometric phase principle, to achieve a continuous phase tailoring over the range of $2\pi$, and at the same time, maintain the desired nonlinear efficiency.

It is to be noted that the previously reported conversion efficiency for third-harmonic generation (THG) in silicon metasurfaces with nonlinear PB phase approach remains very low ($\sim 10^{-9} \frac{1}{W^2}$) compared to the same from the metasurfaces using the Huygens principle (average power conversion efficiency $\sim 10^{-4} \frac{1}{W^2}$).[12,16,31,32] Often, amorphous silicon is used as a material platform for fabricating metasurfaces to investigate THG due to its mature fabrication technology. The nonlinear PB phase principle requires circularly polarized fundamental excitation compared to the linearly polarized excitation in case of nonlinear Huygens metasurfaces. Therefore, there appears a disadvantage for amorphous silicon in the context of nonlinear PB phase control because of its isotropic nature, as THG from an isotropic material under circularly polarized excitation is forbidden.[33] Consequently, the resonator design with amorphous silicon for nonlinear PB phase control requires special attention for achieving a high nonlinear conversion efficiency for THG.[31,34]

In this work, with the goal of achieving a highly efficient nonlinear metasurface together with the PB phase principle, we designed our metasurface based on QBICs that provide high-quality factor modes. Numerically, we achieved a strong third harmonic (TH) signal, mainly originating from the nonlinear enhancement due to the QBIC. The experimental observation of the THG provided a very high nonlinear conversion efficiency ($10^{-4}\frac{1}{W^2}$) under circularly polarized fundamental excitation. Interestingly, nonlinear measurements indicate that the dominant contribution to the TH response arises from a magnetic Mie resonance as confirmed by an analytical model that is based on the principle of coupled anharmonic oscillators. Further, to establish the functional capability of our nonlinear metasurface, we experimentally reconstructed an encoded vortex beam array with different topological charges at the triple frequency with high fidelity, as illustrated in Figure 1a. We applied a design algorithm optimized to overcome the challenges of optical vortex beam generation and demonstrated the successful encoding of information in multiple channels of the TH signal simultaneously.

**Results and discussion**

*Numerical design of nanoresonator and metasurfaces*

In our design, we use a two-dimensional array of cylindrical nanoresonators made of amorphous silicon. The unit cell of the metasurface consists of a cylindrical nanoresonator with an off-centered air hole, which alters the rotational symmetry of the cylinder from $C_\infty$ to $C_1$ as shown in Figure 1a. Note that the $C_1$ rotational symmetry helps to introduce the geometric phase to the system, followed by the selection rule for third-harmonic generation under circularly polarized fundamental illumination[34]. The geometrical parameters of the unit cell were optimized based on the QBIC mode to achieve a high nonlinear conversion efficiency for THG under the fundamental excitation of wavelengths around 1300 nm. For this optimization, we use full-vectorial electromagnetic simulations with the finite-element-method implemented in COMSOL Multiphysics (more details can be found in the Supporting Information). The obtained linear transmission of the metasurface for a period of 664 nm under circularly polarized light is shown in Figure 1b. One can observe a broad resonance dip at a wavelength of 1300 nm with a linewidth of ~70 nm, resembling a magnetic Mie mode, and a sharp peak at a wavelength of 1325 nm with a linewidth of ~8 nm, which corresponds to QBICs. The magnetic field plot at the wavelength of 1300 nm (top inset of Figure 1b) shows the magnetic dipole nature of the resonance located in the x-y plane. Figure 1c illustrates the corresponding electric field plots in the x-z and y-z planes for the same magnetic field in the x-y plane (repeated) at 1300 nm. The electric field forms a vortex in the x-z and y-z planes, which is typical for a magnetic-dipole Mie resonance with in-plane magnetic dipoles, in particular aligned with the x-axis and y-axis.[35] The magnetic dipole mode has the ability to enhance the nonlinear conversion efficiency as they allow for efficient coupling of an electromagnetic field to the nonlinear susceptibility of a material. In contrast, the embedded sharp peak at 1325 nm shows an entirely different mode profile compared to the broad resonance at 1300 nm. At 1325 nm, the electric field plot shows an asymmetric vortex in the x-y plane (bottom inset of Figure 1b), which is enhanced in the vicinity of the air hole. The corresponding magnetic field in the y-z plane is parallel to the cylinder axis, as shown in the lower row of the Figure 1d, along with the same electric field in the x-y plane.

To better understand the QBIC, we consider the nanoresonator without the small air hole. The electric and magnetic field plots of such a cylinder are shown in the upper row of Figure 1d. As visible in the magnetic field plot, this geometry supports a vertical magnetic dipole mode, parallel to the cylinder axis; therefore, the electric field of the mode forms a symmetrical vortex in the x-y plane. The mode is incapable of coupling to the free space as the mode's symmetry in the x-y plane is mismatched to the symmetry of a plane wave at normal incidence, and the unit cell subwavelength period prevents any wavevector which is not normal to the metasurface.[36] These properties eliminate possible channels for radiative decay, resulting in a high-quality factor and a long lifetime. The long lifetime of these so-called BICs is vital for high nonlinear conversion efficiency as it allows for a prolonged interaction between the trapped energy and the nonlinear susceptibility of the material.[25] By incorporating an off-center hole within the cylinder, one can break the in-plane symmetry of the nanoresonator, open radiation channels, and the BIC is then transformed into quasi-bound states in the continuum (QBICs), which is capable of free-space coupling. The corresponding electric and magnetic fields for the QBIC for an array of the cylinders with holes of radius 40 nm are shown in the lower row of the Figure 1d. Compared to the electric field plots for a cylinder without a hole, the hole leaves the electric field vortex mostly intact but results in a field enhancement in its vicinity. This property leads to a nonzero electric dipole moment in the x-y plane, compatible with the symmetry of the plane wave at normal incidence.

As mentioned above, one can use the hole in the cylinder to introduce the Pancharatnam-Berry (PB) phase for circularly polarized light. The qualitative relationship between the rotation $\alpha$ of a nanostructure with $C_1$ rotational symmetry with respect to the laboratory frame and the obtained phase $\phi$ for THG is given by [31,37]:

$$\phi_{Co} = 2\sigma\alpha \quad (1)$$
$$\phi_{Cross} = 4\sigma\alpha \quad (2)$$

where, $\phi_{Co}$ and $\phi_{Cross}$ are the acquired nonlinear phase shifts when the fundamental and TH light has the same (co-), and the opposite (cross-) circular polarization, respectively, and $\sigma = \pm 1$ defines the handedness of the circular polarization state. The principle is

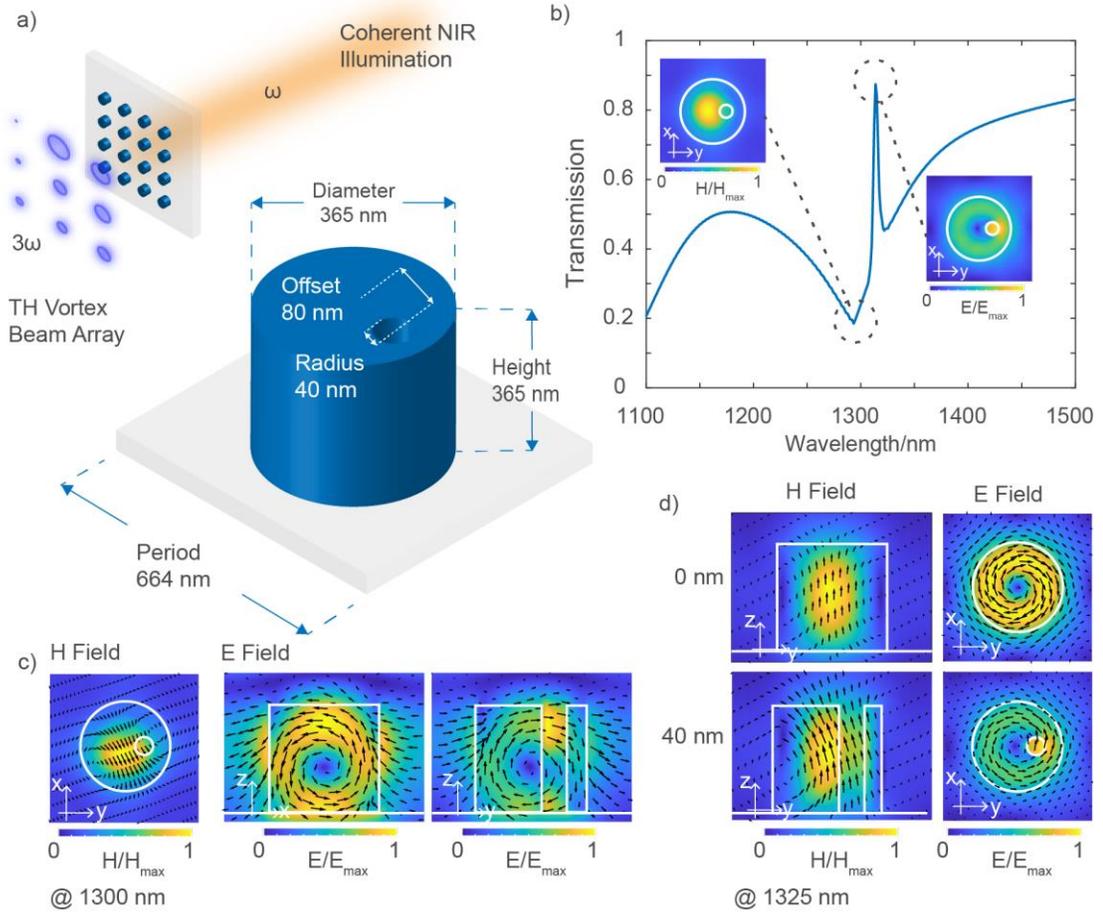

**Figure 1** *a) Unit cell design of a silicon cylinder with broken symmetry. Inset shows the schematic illustration of the projected vortex beams array with different topological charges to be generated at the triple frequency, representing the functional behavior of our metasurface made of silicon cylinders with broken symmetry. b) Transmission spectrum of an array of such cylinders with a hole of radius 40 nm calculated with CST Microwave Studio. The QBICs manifests as a sharp transmission peak inside a broad resonance dip. The insets show the mode profiles for a broad Mie resonance and for the corresponding QBICs. c) Left, x-y plot of the magnetic field (magnetic Mie mode) at 1300 nm inside the amorphous Si cylinder with a hole of radius 40 nm, arranged in a subwavelength array. Right, x-z and y-z plots of the corresponding electric field of the magnetic Mie mode. d) Electric and magnetic field plots inside an amorphous Si cylinder arranged in subwavelength arrays supporting BICs (without the hole, upper row) and QBIC (with a hole of radius 40 nm, lower row). The plots in the x-y plane represent the electric fields at 1325 nm, while the plots in the y-z plane represent the corresponding magnetic fields.*

illustrated in Figure 2a. For functional applications such as in nonlinear holography, a high spatial density of nanoresonators is necessary to achieve high-quality holographic images as the image quality increases with the spatial phase resolution. On the other hand, increasing the density of resonators increases the near-field coupling between the adjacent unit cells[38], leading to a deviation in the TH phase relationship as presented in the Equations 1 and 2. It results in an increased crosstalk between the different polarization channels and introduces a phase noise, limiting the diffraction efficiency. Therefore, due to near-field coupling, there always exists a trade-off between the PB-phase relationship and a high spatial phase resolution

To assess the trade-off between the near-field coupling and the phase, we numerically investigated the TH intensity and the corresponding nonlinear phase in our design for different periods. Figure 2b shows the color plot of the transmission spectra within the wavelength range of 1100-1500 nm as a function of the unit cell period, $p$ (500-900 nm). The transmission spectrum presents a sharp peak within a much broader resonance dip, corresponding to the QBIC as indicated previously in Figure1d. Further, by increasing the unit cell size, the resonance wavelength of the QBIC can be varied from 1225 nm to 1400 nm; however, the peak remains spectrally narrow with a linewidth of ∼8 nm, since the linewidth is mainly controlled by the hole size and its position from the center. However, both parameters were kept constant in the simulations. The plots of the simulated nonlinear phase and TH intensity in the zeroth diffraction order for different unit cell sizes (609 nm to 730 nm) are shown in Figures 2 c-e. It is observed that the TH intensity remains within the same order of magnitude for different unit cell sizes, as expected by the uniform linewidth of the QBIC peak. In addition, the TH intensity in the co-polarization is stronger than the TH intensity in the cross-polarization. Furthermore, the TH signal is highly sensitive to the fundamental wavelength around the QBIC resonance, the intensity of which is strongly reduced when the excitation wavelength differs further from the resonance. The above observations establish QBIC as the main contributor to the THG in the numerical simulation. Figures 2c-e also displays the TH phase change as a function of the rotation of the cylinder at the QBIC wavelength. The rotation angle $\alpha$ of a single nanoresonator between 0° and 180° changes the phase of the generated TH light within 0 to $2\pi$ for co-polarization and within 0 to $4\pi$ for the cross-polarization due to the different dependency of the acquired nonlinear phase on the rotation angle of the nanoresonator ($2\alpha$ and $4\alpha$ in co- and cross-polarizations, respectively). Besides, the nonlinear phase relationship follows higher linearity with larger periods, as the near field coupling becomes weaker with the larger period. Therefore, the unit cell size is crucial for optimizing the system for nonlinear PB phase control, while maintaining a uniform TH intensity.

*Experimental outcomes and analytical modelling*

Based on our design, we fabricated 10 different metasurfaces of size 100x100 µm² for five different unit cell periods (562 nm, 609 nm, 664 nm, 730 nm, 811 nm) and for each period there are two cases, one with and one without a linear spatial PB phase gradient. For the phase gradient, we incrementally increased the rotation angle of the nanoresonator along the x-direction by 22.5°; therefore, a full rotation of 360° can be covered up by 16 unit cells. Two scanning electron microscopy (SEM) images of the metasurfaces, one without a phase gradient and another with a phase gradient, are displayed in Figures 3a and 3b, respectively. Details about the nanofabrication processes can be found in the Method Section.

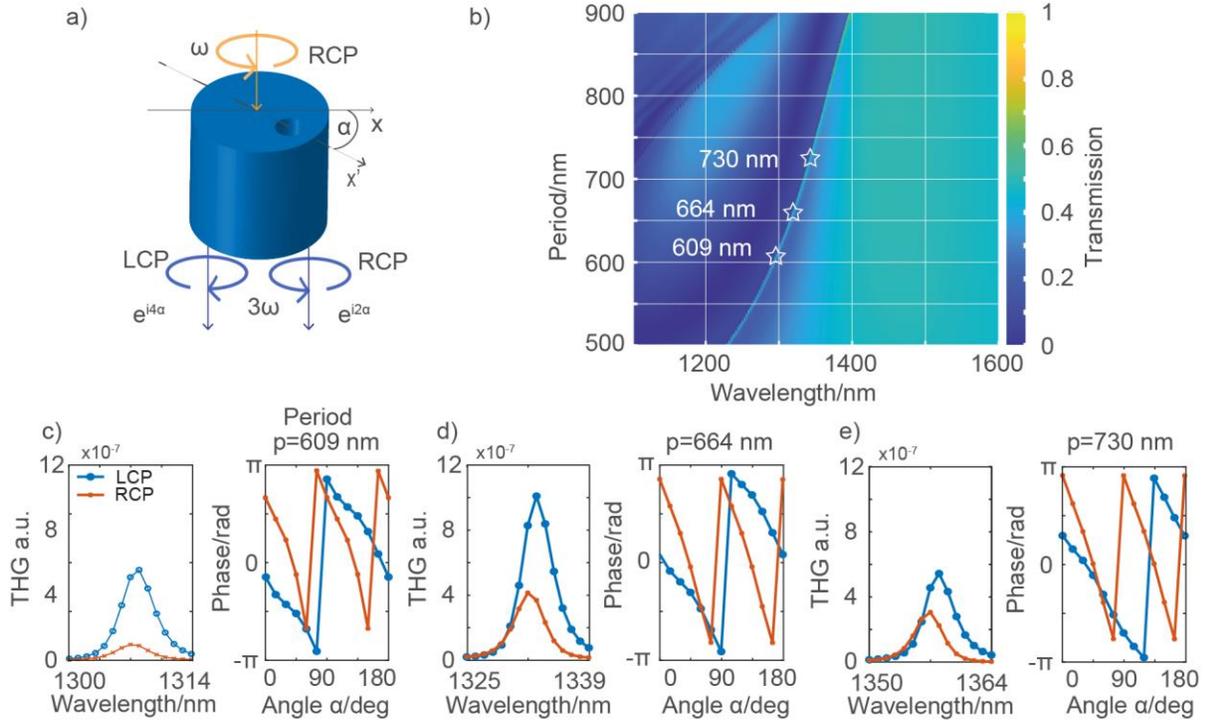

**Figure 2** *a) PB phase principle for THG. A right circularly polarized beam is incident on a cylinder with an off-centered hole ($C_1$ rotational symmetry) which is rotated by an angle α with respect to the laboratory frame. Because of the nonlinear interaction in the cylinder with $C_1$ rotational symmetry, at the TH frequency, the light carries a phase of 2α in the co-polarization (RCP to RCP) and 4α in the cross-polarization (RCP to LCP). b) Two-dimensional plot of the transmission spectra of the metasurface composed of cylinders with an off-centered hole for different periods ranging from 500 nm to 900 nm. The QBICs forms a transmission peak in a broader resonance dip. The insets represent different periods for the nonlinear investigation. c-e) Left: Plots of the TH signal strength as a function of the fundamental wavelength around the QBICs resonance for the three different periods (609 nm, 664 nm, and 730 nm) as marked on b). The fundamental wavelength ranges from 1300 nm to 1314 nm (609 nm period), 1325 nm to 1339 nm (664 nm period) and 1350 nm to1364 nm (730 nm period) which corresponds to a TH wavelength of 433.33 nm to 438.00 nm, 441.67 nm to 446.33 nm and 450.00 nm to 454.67 nm, respectively. Right: Variation of the nonlinear PB phase associated with the generated TH light upon rotation of the $C_1$ cylinder within 0° to 180° for the fundamental wavelengths of 1307 nm (c), 1332 nm (d), and 1357 nm (e).*

Performing a Circle Hough Transformation on several SEM images, we obtained the average radius of the cylinders to be (182.5 ± 3.5) nm, which matches to our design (182.5 nm). However, the hole radius after the fabrication appears to be smaller than the design and increases with the period due to the decrease of the proximity effect during electron beam lithography. Therefore, instead of having a uniform radius of 40 nm, the hole radius in our sample varies from 20 nm for a period of 562 nm to 35 nm for 811 nm.

Figures 3c-d display the plots of the linear transmission spectra for the different unit cell periods for vertically and horizontally polarized incident light, respectively. The polarization states are defined in the inset. Note that the QBIC state is accessible only by the vertically polarized incident light because of the vertical orientation of the net dipole moment of the QBIC. For both polarizations, we observe a broad transmission dip within the wavelength range of 1250 nm to 1300 nm with an apparent difference between vertically and horizontally polarized excitations. The resonance dip for the horizontally polarized excitation lies at a shorter wavelength than the same for the vertical polarization for the same period. In agreement with the simulation, the resonance frequency shifts to longer wavelengths with an increasing period. Moreover, the resonance shape differs for both polarizations; the dips for the horizontal polarization are symmetric, while it appears as asymmetric for vertical polarization, which may result from the influence of the QBIC mode. However, the anticipated peak in the transmission spectra due to QBIC as observed in the simulation (Figure 1d) is not observed experimentally.

Figures 3e-f show the results of the TH measurement from the metasurfaces without phase gradient for different periods for the same and opposite polarizations to the fundamental beams, respectively. The red line in the plots of Figure 3e and 3f represents the TH response of the model of an anharmonic oscillator, which will be discussed later. A similar plot, which shows the same trend for the metasurface with phase gradient, can be found in the Supporting Information. The measurements were made for the fundamental excitations within the range of 1240 nm to 1360 nm. The figures show that the intensity of the generated TH light at the co-polarization is higher than the same at the cross-polarization. Further, the overall TH intensity for co- and cross-polarization is increasing with the unit-cell size. The increase in TH intensity for a longer period

in connection with the decrease in linewidth of the magnetic Mie mode may be associated with the increase of the quality factor. All the nonlinear measurements reveal a broadband TH response of the metasurfaces, accompanied by a peak, representing a few times higher intensity than its surroundings. The spectral width (FWHM) of the TH response is 5 nm to 6 nm. We estimated the maximal achievable nonlinear conversion efficiency of our fabricated metasurfaces. For an average laser power of $100.00 \pm 1.00$ mW ($P_{in}$), which is measured with a power meter, we estimate an average THG power of $0.14 \pm 0.02$ µW ($P_{THG}$), which is derived from the spectrometer data after normalizing to the optical components in the beam path and the quantum efficiency of the detector. It gives the nonlinear peak power conversion efficiency, $\hat{\eta}_{THG} = \frac{\hat{P}_{THG}}{(\hat{P}_{in})^3} \approx 10^{-14} \frac{1}{W^2}$, where the pulse length and repetition rate are of 200 fs and of 80 MHz, respectively, the average power conversion efficiency is $10^{-4} \frac{1}{W^2}$. Note that we set the TH pulse length equal to the pulse length of the fundamental beam because of the low quality factor (Q ~ 40-50) of the resonances and the subwavelength height of the metasurfaces in propagation direction of the fundamental wave.

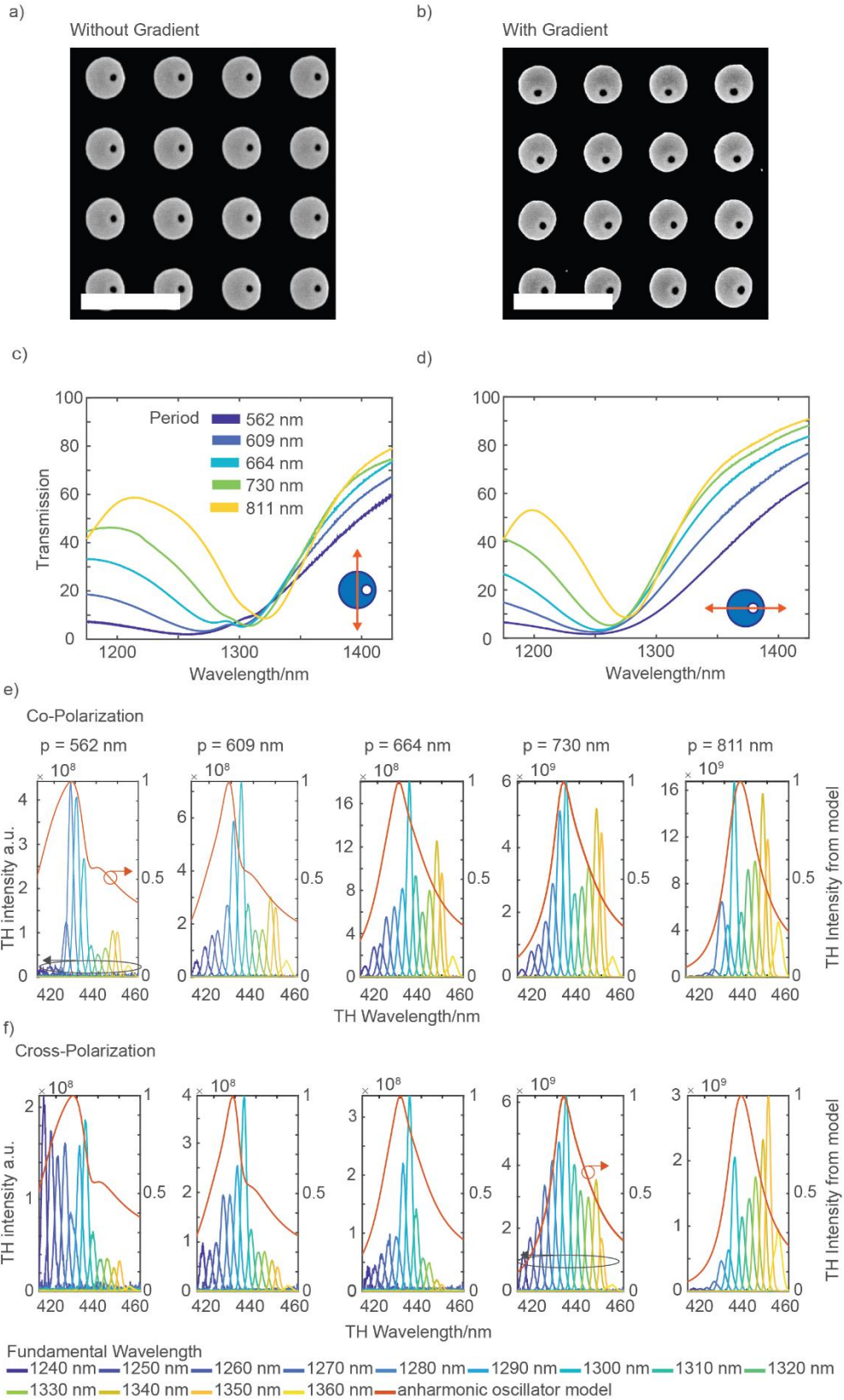

**Figure 3** *a,b) SEM images of etched metasurfaces with and without gradient. Scale bar 1 µm. c,d) Experimentally measured transmission spectra for vertically (c) and horizontally (d) polarized input light of wavelength in the range, 1240-1360 nm for different unit cell sizes. e,f) Intensity of the third harmonic light for the fundamental beam of wavelengths ranging from 1240 nm to 1360 nm for different periods in (e) co- and (f) cross-circularly polarized configurations. The fitted red curves show the overall nonlinear response of an anharmonic oscillator, where the fitting parameters of the model are extracted from the linear transmission spectra in Figure 3c.*

Also note that the estimated efficiency is based only on the zeroth diffraction order of the two-dimensional metagrating. As the TH wavelength is smaller than the period of the unit cell one can expects higher diffraction orders to appear. The higher diffraction orders appear with an angle greater than 45° and thus are not collected by the microscope objective, as the numerical aperture only allows for collecting light within a cone of 24°. Therefore, if we include the 9 diffraction orders (because of the two-dimensional grating), the overall TH conversion efficiency will be one order of magnitude higher. The efficiency remains the same in the case of the corresponding phase gradient metasurface. Compared to an unstructured silicon film of the same thickness, the patterned metasurface shows a 10,000 times higher TH intensity (see Supporting Information). It is worthy to mention that the THG peak power conversion efficiency (in the zeroth order) of the metasurface, when illuminated by a vertically polarized fundamental beam is $\hat{\eta}_{\text{THG,lin}} = \frac{\hat{P}_{\text{THG}}}{(\hat{P}_{\text{in}})^3} \approx 10^{-13} \frac{1}{W^2}$ ($10^{-3} \frac{1}{W^2}$ is the average power conversion efficiency). By using a circularly polarized fundamental beam, we lose 50% of the power during excitation, leading to $\sim \left(\frac{1}{2}\right)^3 = \frac{1}{8}$ times weaker signal, which is almost one order of magnitude lower than for a linearly polarized excitation.

Direct comparison of the linear and nonlinear experimental findings with the corresponding simulations reveals a disparity between the observed results. The simulated linear transmission spectra show a defined peak with narrow linewidth related to the presence of the QBIC while this feature is absent in the experiment. Further, while the nonlinear simulation predicts a strong THG in the vicinity of that peak, our fabricated metasurface presents a broadband nonlinear response in the experiment. For a better assessment of this discrepancy between the experimental and the numerical outcomes, we adopted a model based on a coupled anharmonic oscillator[39]. In this model, we define the amplitude of the nanostructure's magnetic Mie mode as $x_D$ which can interact with an external excitation and couple to the free space. We also define the BIC amplitude as $x_B$ which cannot be coupled directly to the free space but is coupled to $x_D$. Therefore, one can obtain two coupled differential equations as follows:

$$\ddot{x}_D + 2\gamma_D \dot{x}_D + \omega_D^2 x_D - k x_B + \alpha x_D^3 = -f_1 \quad (3)$$

$$\ddot{x}_B + 2\gamma_B \dot{x}_B + \omega_B^2 x_B - k x_D + \alpha x_B^3 = 0 \quad (4)$$

Where $x_{D/B}(t)$ $\gamma_{D/B}$, $\omega_{D/B}$ represent the amplitude, damping, and resonance frequency of the Mie mode (D) and the BICs (B), respectively, and $k$ is the coupling constant between the oscillators. Further, only the Mie resonance is driven by an external electromagnetic field $f_1$, since the BIC is a dark mode. The anharmonic term $\alpha x_{D/B}^3$ gives rise to the nonlinear response of the model, where $\alpha$ is the nonlinear coefficient. The solutions of the coupled equations can be found in the form of a Taylor expansion $x_{D/B}(t) = x_{D/B,0} + \alpha x_{D/B,1} + O(n^2)$. By inserting the Taylor expansion into Equations 3 and 4, the coupled anharmonic oscillator leads to two sets of coupled equations where, $x_{D/B,0}$ is the solution corresponding to the classical coupled harmonic oscillator, while $x_{D/B,1}$ describes the nonlinear response of both oscillators. After solving the displacements $x_{D/B,0}$ the nonlinear response of the coupled oscillators can be calculated as $x_{D/B,1} \sim F[x_{D/B,0}^3]$.[39] The nonlinear response of the metasurfaces can be estimated by using the extracted values of $\gamma_{D/B}$, $\omega_{D/B}$ and $k$ from the fitting of the extinction spectra experimentally obtained from the transmission spectra shown in Figure 3c. The fit values can be found in Table 1 and details about the fitting procedure can be found in the Supporting Information: Figures 3e-f show the comparison between the experimentally observed TH response for a wavelength range for co- and cross-polarized configurations and the corresponding TH response from the model calculation, $I_{THG} = |x_{D,1} + x_{B,1}|^2$ (red curve). The model can reproduce the experimentally observed broadband nonlinear response qualitatively by only considering the linear optical properties of the system. However, from the quantitative point of view the model may not provide a careful estimation as it does not include transmission and reflection from the oscillator but only the oscillator strength and the model parameters are derived from the measured linear transmission results. Furthermore, the model describes the metasurfaces as a single coupled oscillator. Therefore, small geometric changes can have a larger impact on the nonlinear optical properties, while the linear optical properties are mostly remain unchanged. Additionally, the nonlinear susceptibility of the material is also not considered in this model but only the nonlinear behavior based on the resonance itself, which also influence the deviation between the model and the experimental data. Besides, the coupled anharmonic oscillator model, temporal coupled mode theory (CMT) could also be useful to describe and analyze the third harmonic response from resonant nonlinear systems[40]. However, our model reasonably predicts the overall nonlinear response of the metasurface, and we can separate the individual contributions from the two individual modes. Figure 4a shows the overall THG response from the model and the individual contributions from $x_{D,1}$ and $x_{B,1}$ exemplary for the metasurface with a period of 664 nm. Interestingly, the calculated nonlinear response from the model is dominated by the nonlinear oscillator strength $x_{D,1}$ that corresponds to the magnetic Mie mode, while the contribution from the BIC, $x_{B,1}$ is small, since the pump mainly excites the Mie mode. Note that the resonance wavelengths, extracted from fit, associated with the frequencies $\omega_D$ and $\omega_B$ of the magnetic Mie mode and the QBIC fall into the broad transmission dip (1287 nm), and the asymmetric shoulder (1293 nm) in the transmission spectrum corresponds to a period of 664 nm. The linewidths of the dipole mode and BIC are estimated as 52 nm and 12 nm, respectively. The same linewidths can also be estimated from the fitting of the other experimental transmission spectra associated with different other periods, and in every case, the nonlinear oscillator strength $x_{D,1}$ dominates the overall nonlinear response.

**Table 1** *Model parameters extracted by fitting the numerically and experimentally obtained transmission spectra through equations 3 and 4. Note that the values in the table are converted to wavelength for better comparison with Figures 1b and 3c.*

|  | $p$ [nm] | $\omega_d$ [nm] | $\gamma_d$ [nm] | $\omega_B$ [nm] | $\gamma_b$ [nm] | $k$ [nm] |
|---|---|---|---|---|---|---|
| Experimental spectra (Figure 3c) | 563 | 1263 | 49 | 1307 | 12 | 19 |
|  | 609 | 1276 | 42 | 1296 | 11 | 22 |
|  | 664 | 1290 | 39 | 1294 | 11 | 21 |
|  | 730 | 1301 | 32 | 1296 | 19 | 20 |

|  | 811 | 1313 | 39 | 1315 | 23 | 32 |
| --- | --- | --- | --- | --- | --- | --- |
| Simulation (Figure 1b) | 664 | 1300 | 25 | 1320 | 0.1 | 16 |

For further investigation, we fit our model to the numerical transmission spectra as displayed in Figure 1b. The fit values can be found in Table 1. The resonance wavelengths corresponding to the frequencies $\omega_D$ and $\omega_B$ of the magnetic Mie mode and the QBIC fall at the minimum of the broad dip and the sharp peak in the transmission spectra, respectively. In contrast to the linewidths for BIC and Mie modes observed in the experiment, the linewidths extracted from the simulation are appeared to be smaller. The fitting presented in Table 1 reveals a linewidth of ~40 nm for the Mie mode, which is close to the value obtained from the experimentally measured transmission spectra. However, in Table 1 the linewidth $\gamma_b$ of the mode $\omega_B$ is estimated as 0.1 nm, which corresponds to a quality factor of 13.200, agrees with the assumption that the dark mode $x_B$ in this resonator system resembles a BIC mode. In addition, the coupling constant can be estimated as $k = 16$ nm, which is smaller than the linewidth of the magnetic Mie mode, indicates a connection with the Fano resonant regime associated with weak coupling[41]. Therefore, in the numerical case, our design supports BIC and the coupled anharmonic oscillator established by equation 3 and 4 describes a QBIC for the set of parameters presented in Table 1. Figure 4b shows the nonlinear response of the model fitted to the simulated transmission spectra, where the BIC ($x_{B,1}$) mainly contributed to the nonlinear response. However, for a slight change of the linewidth of the BIC, the TH response is largely reduced and becomes broadened, and the contribution of the dipole oscillator becomes dominant. Note that the total TH intensity is larger than the sum of the contributions from the individual oscillators as square of the sum of the two is larger compared to the sum of the two individual squares. In the context of this model, compared to the simulation, the linewidth of the BIC is broadened by several orders of magnitude in the experiment, as presented in Table 1. Therefore, the linewidth $\gamma_B$ of the QBIC is the key parameter that makes our experimental observations significantly different from the numerical results, the linewidth of the QBIC appears to be around 12 nm in the experiment. Further, the model parameters can be extracted from the simulated transmission spectrum shown in Figure 1d and the linewidth $\gamma_B$ can be increased from the original value of 0.1 nm to 6.5 nm, which is close to the values extracted from the experimentally obtained transmission. By increasing the linewidth $\gamma_B$ in the anharmonic oscillator model and keeping the remaining model parameters constant, the TH signal strength of $x_{B,1}$ decreases and $x_{D,1}$ becomes the dominant contributor to the overall nonlinear oscillator strength, as shown in the Supporting Information. Moreover, the transmission spike which is visible in Figure 1d becomes less pronounced and resembles the measured transmission in Figure 3c, as the linewidth is increased from 0.1 nm to 6.5 nm. Besides the linewidth, the spectral distance of the BICs and Mie resonance appears to influence the nonlinear response of the system as well. In total, the dominant contribution in the experiment stems from the magnetic Mie mode and not from the QBIC since the linewidth of the QBIC, which represents the loss of this mode, appeared to be too large to drive the QBIC effectively. Reason for the large losses, which limit the contribution of the QBIC mode can be mostly fabrication related. This includes geometrical variations between the resonators during the lithographic patterning, as explained in Ref.[42], and changes in the refractive index during the reactive ion etching and the other fabrication steps. In addition, potential surface roughness of the resonators, which also leads to a broadening, was not included in our design. Overall, these parameters can all limit the quality factor of the mode. Note that QBIC in silicon metasurfaces with high quality factors were demonstrated.[43,44]

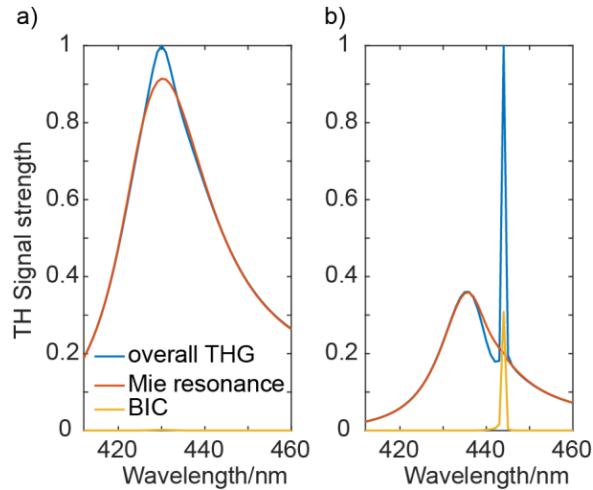

**Figure 4** *Plots of the TH signal strength as estimated from the anharmonic oscillator model by utilizing the model parameters extracted by fitting a) the experimentally measured linear transmission spectrum for a period of 664 nm, displayed in Figure 3c, and b) the simulated spectrum for the same period as displayed in Figure 1b. The overall TH response $I_{THG} = |x_{D,1} + x_{B,1}|^2$; $I_D = |x_{D,1}|^2$ and $I_B = |x_{B,1}|^2$ are the contributions from the magnetic Mie mode and the BICs, respectively. The plots are normalized to the maximum oscillator strength of $I_{THG}$.*

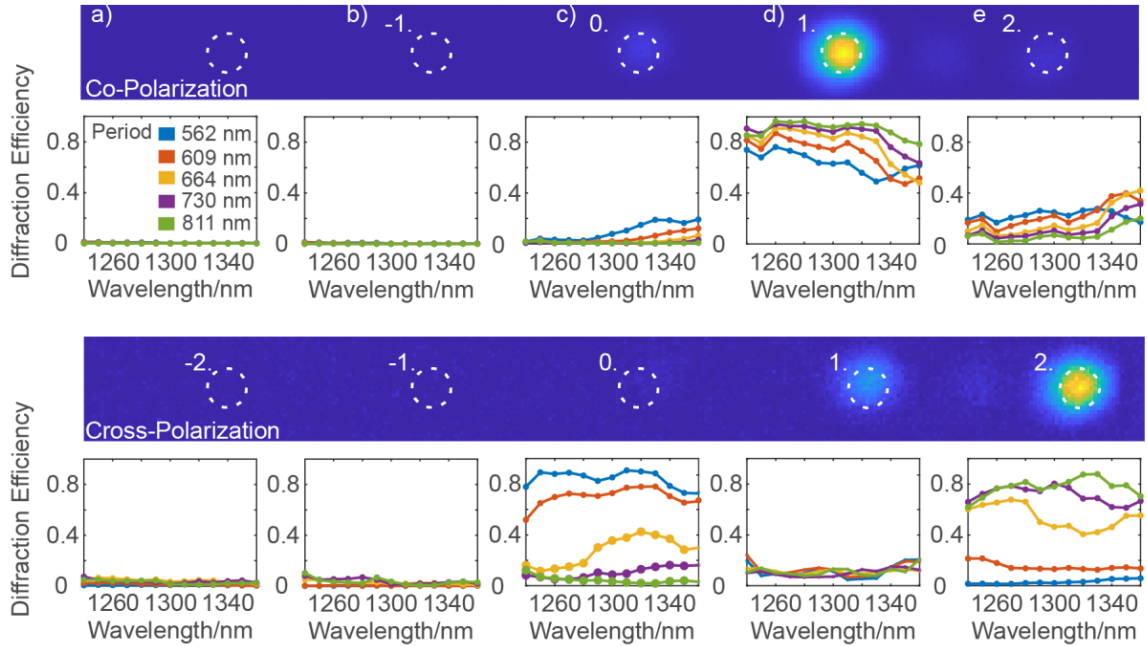

**Figure 5** *Diffraction efficiencies of the TH light in different diffraction orders for different periods of the phase gradient metasurface under the excitation of the fundamental beams with wavelengths ranging from 1240 nm to 1360 nm in co- and cross-circularly polarized configurations. The Fourier images, above the plots of the diffraction efficiencies, are illustrated for a period of 811 nm and for a fundamental wavelength of 1300 nm. The first row shows the same for co-polarized measurement, while the second row for the cross-polarized measurement.*

To realize the nonlinear phase-tailoring properties experimentally, we investigated THG from our fabricated metasurfaces with the desired phase gradient based on the PB phase principle for different unit cell periods, leading to a beam deflection (for details, see Supporting Information). Depending on the period, one expects diffraction orders at around 4.64° (co-polarization) and 9.29° (cross-polarization) for a TH wavelength of 430 nm. As the phase gradient resembles a blazed grating, the diffraction spots can be assigned to different phase gradients resulting from the phase factors of the PB-phase ($\pm 2^{nd} \leftrightarrow \pm 4\alpha$, $\pm 1^{st} \leftrightarrow \pm 2\alpha$), as explained in the Supporting Information. We spatially resolve the TH signal in the Fourier-space by imaging the back focal plane of the microscope objective that collects the light from the metasurface. The measured TH intensity in different diffraction orders, along with the estimated diffraction efficiencies for co- and cross-polarization, are shown in Figure 5. The k-space image for the co-polarized TH light shows a prominent spot at $1^{st}$ diffraction order (at the desired angle) with unwanted weaker contributions at the $0^{th}$ and $2^{nd}$ order location. In contrast, the cross-polarized TH light appears mainly in the $2^{nd}$ diffraction order. The $0^{th}$ order diffraction spot in the center of the Fourier space does not carry any PB phase. We calculated the diffraction efficiency for a particular diffraction order as the quotient of the intensity of that diffraction order (marked by the dotted white lines) to the overall TH intensity.

In co-polarization, the first diffraction order (proportional to $2\alpha$) appears with very high diffraction efficiency (~90%) for periods ranging from 664 nm to 811 nm. The high diffraction efficiency is maintained for a broad range of fundamental wavelengths. However, for smaller periods (609 nm and 562 nm), the diffraction efficiency decreases dramatically due to the upsurge of the near-field coupling strength between the adjacent nanoresonators. Moreover, the diffraction efficiency for the $-1^{st}$ (Figure 5b) and $-2^{nd}$ (Figure 5a) diffraction orders are negligible as the phase gradient resembles blazed gratings. The same trend of variation of the diffraction efficiency is also observed for cross-polarized detection configuration. For cross-polarization, the phase shift is proportional to $4\alpha$ and therefore, the maximum diffraction efficiency occurs at the $+2^{nd}$ diffraction order (Figure 5e). Here, we achieve a maximum diffraction efficiency of ~80%. However, the overall diffraction efficiency is estimated as lower for the cross-polarization compared to the co-polarized configuration. This is primarily because the TH phase in the cross-polarization changes with $4\alpha$ while in the co-polarization case the phase changes with $2\alpha$, decreasing the phase resolution by a factor of 2. Consequently, the diffraction efficiency decreases for the cross-polarization compared to the co-polarization.

To establish the functional capacity of our metasurface, we designed a vortex beam array for its realization at the triple frequency. The design was based on the spatial distribution of the TH phase to accomplish the desired nonlinear diffractive meta-element. We chose a 3x3 arrangement of vortices with varying topological charges $m$ for the conversion of an original Gaussian beam at the fundamental wavelength into multiple vortex beam profiles at the TH wavelength. By considering the trade-off between the conversion and diffraction efficiencies, we designed and fabricated the metasurface with a period of 664 nm (for details of the design, see Supporting Information). By utilizing the design freedom of the PB-phase principle, we encoded two different vortex beam arrays in the co- and cross-polarization channels of the TH light (corresponding to the phase factors $2\alpha$ and $4\alpha$, respectively). The obtained experimental and numerical results for the nonlinear optical vortex array generation with respective topological charges are shown in Figure 6. The small spot in the center of the Fourier plane, visible in the experimentally reconstructed arrays, arises from residual TH light, which is not carrying any PB-phase information. Note that the vortex beams with higher topological charges start to interfere with each other (e.g., the vortex beams with topological charges 14 and 16, shown in Figure 6c and 6d) since the vortices are densely packed in an angular range of 17.46° (NA = 0.3). For a careful observation, one can see that the measured vortex arrays exhibit a minor speckle pattern, which is typical for phase-only holograms. By inverting the polarization state of the fundamental illumination, the patterns generate their conjugated images due to the sign change of σ in the phase factor of the THG signal. Besides, a different vortex beam array was encoded at the fundamental wavelength by means of wavelength multiplexing. The reconstructed image in the cross-polarization channel as determined by the linear PB phase principle (only allowed in cross-polarization), at the fundamental wavelength can be found in the Figure 6e and f. The vortex beam array is reconstructed within an angular range of 36.86° (NA = 0.6), therefore interference between different vortices is small. Again, by inverting the polarization state, the conjugated vortex beam can be reconstructed. Careful observation reveals that there appear ghosts images associated with the individual vortex beams in an array as a result of the internal reflections inside the camera, since the anti-reflection coating and the camera itself are optimized for the visible spectral range. This effect is stronger for the vortex beam with low topological charges, e.g., for m = -2, as the intensity is concentrated over the smaller area on the camera sensor.

**Conclusion**

In this work, we demonstrate both numerically and experimentally a nonlinear silicon metasurface that can generate third harmonic (TH) light very efficiently and simultaneously can tailor the nonlinear phase based on the Pancharatnam-Berry (PB) phase principle. Our metasurface experimentally presented a strong broadband third harmonic response with an average power nonlinear conversion efficiency of $\eta_{\text{THG}} = \frac{P_{\text{THG}}}{(P_{\text{in}})^3} \approx 10^{-4} \frac{1}{W^2}$ when illuminated with a circularly polarized fundamental near-infrared laser beam (peak power conversion efficiency, $\hat{\eta}_{\text{THG}} = \frac{\hat{P}_{\text{THG}}}{(\hat{P}_{\text{in}})^3} \approx 10^{-14} \frac{1}{W^2}$). The metasurface was designed to support optical quasi-bound states in the continuum modes, however, experimentally the magnetic Mie mode dominates the overall TH response with a very weak contribution from the QBICs, as confirmed by an analytical model based of an anharmonic oscillator. Besides, our metasurface provides a high diffraction efficiency of ~90% in co-polarization and ~80% in cross-polarization at the third harmonic wavelength. The conventional designs of the nonlinear phase control with all-dielectric metasurfaces reported in literature are based on the nonlinear extension of the Huygens principle, which demands a large set of nanoresonators with different geometries to achieve a complete $2\pi$ phase coverage at a particular single frequency. Furthermore, it requires huge computational resources in the design.[13,16] In contrast, the nonlinear extension of the PB phase principle, which requires a more straightforward design and a better-simplified nanofabrication process, can end up with very high conversion and diffraction efficiencies even at a circularly polarized fundamental excitation, along with a continuous and full control of the nonlinear phase, as evidenced by this work. Finally, we established the functional capability of our metasurface by illustrating the experimental reconstruction of a polarization multiplexed encoded vortex beam array supported by varying topological charges, with high fidelity and high conversion efficiency.

The concept provides a pathway to realize the miniaturized control of the high conversion and diffraction efficiencies to be applicable in nonlinear nanophotonic devices where advanced phase control with low spatial footprints is necessary, such as nonlinear optical information processing and wavefront engineering.

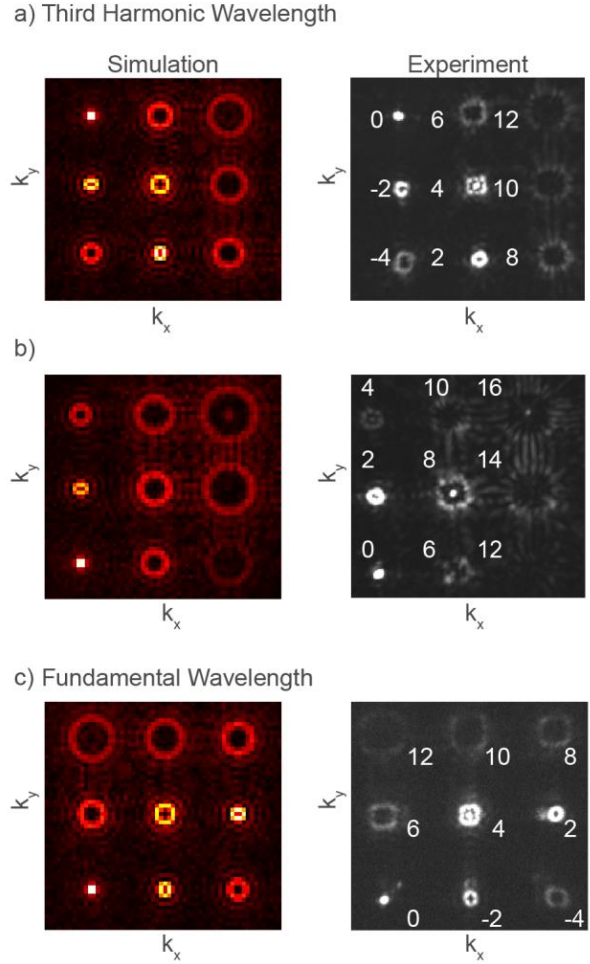

**Figure 6** *The left column shows the numerically simulated vortex beam arrays of varying topological charges at the TH wavelength in a) co-polarization, and in b) cross-polarization. The right column displays the experimentally reconstructed TH images at the Fourier plane carrying the phase of the same vortex beam arrays in a) co-polarization, and in b) cross-polarization. The numbers on the experimentally reconstructed figures represent the topological charges of each generated vortex beam. c) Exhibits the numerically simulated (left) and the experimentally (right) measured vortex beam arrays in cross-polarization, respectively, at the fundamental wavelength.*

**Materials and methods**

*Optical experiments*

The linear transmission spectra were measured using a white light Laser source (Fianium Whitelase) in the spectral range of 1150 nm to 1300 nm. Further, we controlled the input polarization by a linear polarizer. A 10x infinity-corrected microscope objective collected the metasurface's transmission. The light was routed to a spectrometer (Andor Shamrock Kymera 193) with a set of lenses. The spectrometer was equipped with an InGaAs detector (Andor IDus 491A InGaAs) suitable for measurements in the infrared. The spectral resolution of the transmission setup is 2.43 nm.

We use an Optical Parametric Oscillator with a pulse length of 200 fs and an 80 MHz repetition rate as a laser source for the nonlinear optical experiments. We change the wavelength between 1200 nm and 1350 nm during the experiment but keeping the average laser power at 50 mW. The laser light is focused on the metasurface by a 50 mm achromatic lens optimized for the infrared resulting in a beam waist size of approximately 45 µm. Therefore, the average pump peak power density of $0.32 \frac{GW}{cm^2}$.

To collect the THG light from the metasurface, we use a 50x microscope objective with a numerical aperture of 0.42. By a set of two lenses, we capture the back focal plane into an imaging spectrometer. The imaging spectrometer (Andor Shamrock 303 equipped with an iDUS 420 detector) allows us to resolve the frequency-converted light's wavelength and different diffraction orders orthogonal to the diffraction grating. Using a combination of linear polarizers and quarter-wave plates, we prepare and analyze the circular polarization state. We have used short pass filters to suppress the fundamental Laser beam, to minimize nonlinear contributions from other optical components. The holographic images were captured with a similar setup as discussed above, but an sCMOS camera (Andor Zyla 4.2) replaces the spectrometer to capture a high-fidelity image of the whole Fourier space.

*Vortex beam array design*

The vortex beam array metasurface design exploits the same QBIC metasurface as discussed above, but a computer-generated holographic (CGH) algorithm determined every nanoresonator's rotation on the metasurface. This algorithm is optimized for vortex beam generation at the Fourier plane. There is a particular need for optimized algorithms since the generation of optical vortex beam arrays has several challenges, such as low quality and an unfavorable energy distribution among the vortex beams. Further, the algorithm allows for polarization and wavelength multiplexing, satisfying the demand for compact, highly functional meta-devices.

To establish the functional capacity of our nonlinear metasurface, we designed a nonlinear Dammann vortex beam arrays for its realization at two circular polarization channels of THG and third channel in linear case. Since for such C1 symmetric meta-atom, one can predict $2\sigma\theta$ ($\sigma = -1$) for opposite handedness at the fundamental frequency, and $2\sigma\theta$ ($\sigma = 1$), $4\sigma\theta$ ($\sigma = -1$) for same and opposite handedness at the THG frequency, respectively. The transmission function can be expressed by Fourier series as follows:

$$T_{NDVG}^{(n\pm1)\sigma\theta} = \exp(i(n\pm1)L_0\theta) \times \sum_{m=-\infty}^{\infty}\sum_{n=-\infty}^{\infty} T_{mn}^{(n\pm1)\sigma\theta} \exp\left(i\frac{2\pi}{T_x}(mx+ny) + i(mL_x + nL_y)\theta\right) \quad (5)$$

Where the inset topological charges are set as $L_0 = 2$, $L_x = 2$, $L_y = 6$, n represents the harmonic generation order, which can be 1 (for fundamental frequency) and 3 (for THG). While (n-1) and (n+1) correspond to the same handedness and opposite handedness of the output light with respect to the incident one, respectively. $C_{mn,2\sigma\theta}$ and $C_{mn,4\sigma\theta}$ are the optimizing target to get desired diffraction orders as 3x3. As demonstrated, the topological charges of each diffraction order obey the rule of $L_0 + mL_x + nL_y$ for $2\sigma\theta$ case and $2L_0 + mL_x + nL_y$ for $4\sigma\theta$ case. For each diffraction order, the numerical aperture NA $= \frac{\lambda}{T_x}\sqrt{m^2+n^2}$, which changes with the generated frequencies. The first item in the series expansion predict the location of each diffraction order, which also relates to the generated frequency. Afterwards, by using genetic algorithm method, the optimized rotation angles of the nanoholes within the cylinders can simultaneously obtain the ideal target orders in both linear and nonlinear channels.

**Data availability**

The data that support the findings of this study are available from the corresponding author upon reasonable request.

**Acknowledgment**

This project has received funding from the European Research Council (ERC) under the European Union's Horizon 2020 research and innovation programme (grant agreement No 724306) and the Deutsche Forschungsgemeinschaft (DFG, German Research Foundation) – SFB-Geschäftszeichen TRR142/2-2020 – Projektnummer 231447078 – Teilprojekt C05. We also acknowledge the NSFC-DFG joint program (DFG No. ZE953/11-1, NSFC No. 61861136010) for continuous support.